\documentclass[fonts]{icst}
\usepackage{moreverb}

\usepackage{url}
\usepackage[breaklinks,colorlinks,bookmarksopen,bookmarksnumbered,linkcolor=ICSTblue,citecolor=blue,urlcolor=ICSTblue]{hyperref}

\usepackage{diagbox}
\usepackage{siunitx}
\DeclareSIUnit \dBm {dBm}
\DeclareSIUnit \dB {dB} 
\DeclareSIUnit \Kbps {Kbps}
\DeclareSIUnit \Mbps {Mbps}
\DeclareSIUnit \Gbps {Gbps}
\DeclareSIUnit \kBps {kBps}
\DeclareSIUnit \MBps {MBps}
\DeclareSIUnit \GBps {GBps}    
\usepackage[caption=false, font=footnotesize]{subfig}
\usepackage{amsfonts}
\usepackage{amsmath}
\usepackage{times}
\usepackage{algpseudocode}
\usepackage{algorithm}
\usepackage{enumerate}
\usepackage{multirow}
\usepackage{tikz}
\usepackage{slashbox}
\usepackage{array}
\usepackage[originalparameters]{ragged2e}
\newcolumntype{P}[1]{>{\centering\arraybackslash}p{#1}}
\newcolumntype{M}[1]{>{\centering\arraybackslash}m{#1}}
\newcolumntype{x}[1]{>{\centering\let\newline\\\arraybackslash\hspace{0pt}}p{#1}}

\newcommand\BibTeX{{\rmfamily B\kern-.05em \textsc{i\kern-.025em b}\kern-.08em
T\kern-.1667em\lower.7ex\hbox{E}\kern-.125emX}}

\def\volumeyear{2018}

\journalname{Industrial Networks and Intelligent Systems}
\articletype{Research Article}
 \def\volumeyear{2018}
 \def\volumenumber{XX}
  \def\issuemonth{XX-XX}
  \def\issuenumber{X}
  \def\articleid{XX}
  \def\copyrightholder{Ioannis Mavromatis\textit{ et al.}, licensed to ICST}
  \copyrightnote{This is an open access article distributed under the terms of the Creative Commons Attribution license (\url{http://creativecommons.org/licenses/by/3.0/}), which permits unlimited use, distribution and reproduction in any medium so long as the original work is properly cited.}
  \def\articledoi{10.4108/XX.X.X.XX}
\received{XXXX}
  \accepted{XXXX}
  \published{XXXX}

\begin{document}

\runningheads{I. Mavromatis, A. Tassi, G. Rigazzi, R. J. Piechocki, A. Nix}{Multi-Radio 5G Architecture for Connected and Autonomous Vehicles: Application and Design Insights}

\title{
Multi-Radio 5G Architecture for Connected and Autonomous Vehicles: Application and Design Insights
}

\author{Ioannis Mavromatis\fnoteref{1}, Andrea Tassi\fnoteref{1}, Giovanni Rigazzi, Robert J. Piechocki, Andrew Nix}

\address{Department of Electrical and Electronic Engineering, University of Bristol, Bristol, UK}

\abstract{Connected and Autonomous Vehicles (CAVs) will play a crucial role in next-generation Cooperative Intelligent Transportation Systems (C-ITSs). Not only is the information exchange fundamental to improve road safety and efficiency, but it also paves the way to a wide spectrum of advanced ITS applications enhancing efficiency, mobility and accessibility. Highly dynamic network topologies and unpredictable wireless channel conditions entail numerous design challenges and open questions. In this paper, we address the beneficial interactions between CAVs and an ITS and propose a novel architecture design paradigm. Our solution can accommodate multi-layer applications over multiple Radio Access Technologies (RATs) and provide a smart configuration interface for enhancing the performance of each RAT.}



\fnotetext[1]{Corresponding authors: I. Mavromatis (\email{ioan.mavromatis@bristol.ac.uk}) and A. Tassi (\email{a.tassi@bristol.ac.uk}).}

\maketitle

\section{Introduction}
The introduction of autonomous vehicles will represent the biggest revolution on our roads since the advent of internal combustion engine. The benefits include traffic reduction and increased traffic predictability, better road safety, new mobility options and social inclusion.    
Recent forecasts estimate that globally the number of people living in urban areas is due to increase to over $66$\% by 2050~\cite{UN}. In particular, road congestion determines substantial productivity losses.
Consider the simple act of searching for a parking space; this represents around $30$\% of all road traffic in mid-to-large cities, mainly due to two factors:
\begin{itemize}
\item a \emph{lack of knowledge} - If the drivers were aware of the presence of a traffic jam, they would try and avoid congested roads by selecting a different route (vehicle rerouting), or they would choose alternative transportation means (road off-loading);
\item a \emph{lack of confidence} - Drivers are not aware of the location of the next available parking slot and cannot quickly reach their destinations, wasting time in painstakingly looking for available spaces and even driving erratically.
\end{itemize}
Traditional Intelligent Transportation Systems (ITS) should allow a user: (i) to plan the journey ahead, and (ii) to react to traffic jams by rerouting vehicles. However, given the expected population growth, road congestion needs to be prevented more proactively.

\begin{figure*}[t]
\centering
\includegraphics[width=0.95\textwidth]{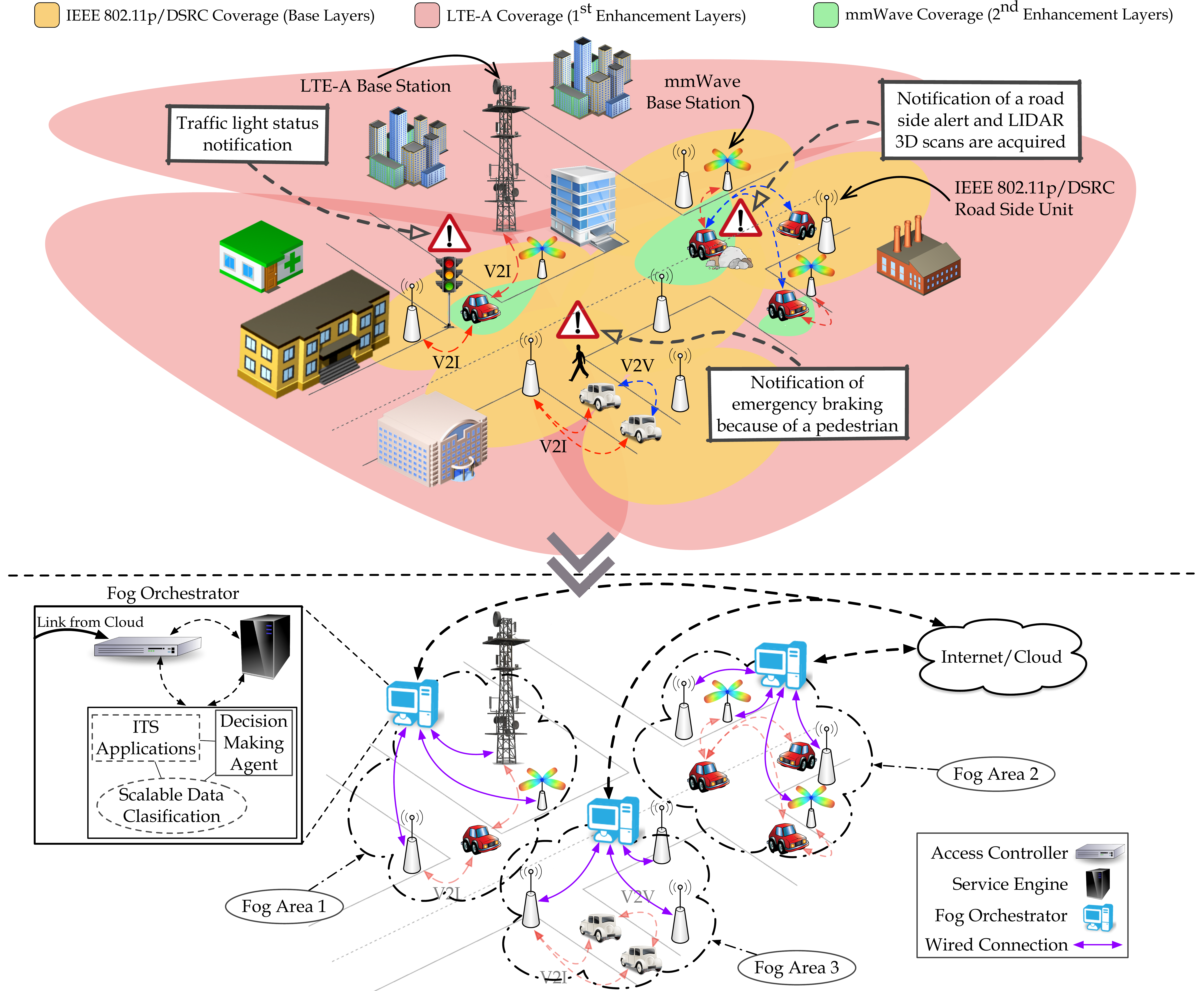}
\caption{General overview of the considered system model. The proposed ITS design framework ensures analog-like performance degradation communications among CAVs by means of multi-layer V2X communications over an heterogeneous network infrastructure.}
\label{fig:NW_T}
\end{figure*}

Next-generation Cooperative ITSs (C-ITSs) are expected to bring the paradigm of Mobility-as-a-Service (MaaS) to a whole new level by means of Connected Autonomous Vehicles (CAVs). A critical factor in a CAV-based MaaS paradigm is represented by autonomous vehicles that cease to be \emph{autonomous systems} and become \emph{cooperative entities}. Specifically, cooperation among autonomous vehicles is enabled by the sharing of sensor data and maneuvering intentions in Vehicle-to-Vehicle (V2V) and Vehicle-to-Infrastructure (V2I) fashion. 
For these reasons, MaaS city models based on CAVs have the potential to overcome the users' lack of knowledge and confidence, by means of intelligent route planning systems, and efficient resource allocation strategies for road systems.

In Section~\ref{sec:DM}, we introduce a novel C-ITS framework based on a heterogeneous communication infrastructure, as shown in Fig.~\ref{fig:NW_T}. Section~\ref{sec.comms} shows how the proposed C-ITS framework enables efficient Vehicle-to-Everything (V2X) communications among CAVs by: i) handling scalable ITS applications consisting of independent data streams, and mapping onto different V2X layers, and ii) employing multiple Radio Access Technologies (RATs) to deliver each data stream, according to target Quality-of-Service (QoS) constraints. In Section~\ref{sec.ITS}, we first illustrate the key components of CAVs and identify the necessity for an abstract plane responsible for the management and coordination of the system. Then, an overview of the candidate RATs is provided, along with a discussion on their benefits and limitations. Finally, we present our envisioned C-ITS architecture and examine all its components as well as discuss prospective V2X use cases. Conclusions and future research avenues are drawn in Section~\ref{sec.cl}.

\section{C-ITS Communication Systems}\label{sec:DM}
The principle of MaaS makes possible for the end-users to access a range of next-generation transportation services aided by accurate navigation, journey information with vastly improved road safety. These services will improve the users' travel experience and provide higher efficiency on how an ITS works. Undoubtedly, a reliable communication system is an enabler for cooperative MaaS frameworks and a key component in next-generation C-ITSs.

CAVs will be equipped with several sensors to maneuver and plan their trajectory ahead. The short-term trajectory planning is essentially achieved by the information gathered from the on-board sensors~\cite{local_planning}. For instance, this is the case of an obstacle (e.g., a pedestrian) avoidance maneuver to be performed~\cite{pedestrian_avoidance}. However, for more sophisticated scenarios (e.g. a CAV intending to park at an empty space beside the next roadblock), the long-term trajectory planning requires two key elements. At first, information from the surrounding environment should be collected~\cite{connected_path_planning}. Secondly, a \emph{Decision-Making agent} is required~\cite{GPMN16}. This agent, using the acquired data, is responsible for the decision-making (e.g. changing lane for better positioning on the road~\cite{lane_changing}, rerouting CAVs due to traffic congestion, etc.). 

Each ITS service has specific requirements regarding the data that should be acquired and exchanged. For example, a ``collision avoidance by cooperative maneuvering'' application is benefited by the almost instantaneous exchange of high-resolution LiDAR or camcorder data~\cite{cam_coder}. Considering that a 3D-LIDAR can generate a data stream of several \emph{hundreds of megabit-per-second}, both the data rate and latency of communication links are fundamental key performance indicators (KPIs) in delivering a high precision and low-risk maneuver~\cite{KSAGKR12}. On the other hand, information such as the position of the next traffic light or the next parking space is of less significance, require lower data rate and is more tolerant to communication latency.

Based on the unique application requirements and taking into account the diverse features of each potential RAT, we designed our system based on a \emph{scalable data architecture}. The idea of scalability in data has originally been applied to video transmissions where a scalable video stream consists of a \emph{base layer} and multiple \emph{enhancement layers}. The base layer allows users to achieve a basic reconstruction quality, which is gradually improved as soon as the enhancement layers are successfully received. The same principle trespassed the natural boundaries of multimedia communications and is being applied to design systems capable of an analog-like service degradation~\cite{Ai}.

The different data streams in our system are mapped onto different RATs and the central coordination takes place in a Software-Defined Networking (SDN)-like approach (Fig.~\ref{fig:NW_T}). The \emph{Decision-Making agent} is responsible for the automated driving applications, which require nearly instantaneous decision-making and reliable communication links. The design of this agent is outside the scope of this work but the reader can go into more details on the topic in~\cite{GPMN16}. 

The separation of the data and the control plane enables two prominent features of this system. At first, it enables the management capabilities and the dynamic network resource allocation required by a fifth-generation (5G) vehicular communication framework~\cite{6834762}. Decisions are being taken with respect to various network KPIs, such as the availability of a RAT or the load of a communication link. Later, a centralized control over the heterogeneous network provides a global view and a unified configuration interface. In our system, we assume that our network is clustered in different management areas called \emph{Fog Areas}. Each Fog Area is centrally managed by a \emph{Fog Orchestrator} (FOs) and consists of a number of multi-RAT Road Side Units (RSUs) and controls multiple CAVs. The FOs represents the logical entities encapsulating the core components of our system. Within these entities, the automated driving decisions are taken and data streams are mapped onto the different RATs.

Utilizing this Fog Computing implementation, our centralized controller will be responsible for the generation and the processing of all the messages. The different radios will be solely responsible for relaying the message to and from the CAVs. Our envisaged solution still interacts with a cloud-based city-wide connection, which it interfaces with the \emph{Access Controller} in the FO. In particular, the cloud-based service will only be in charge of recording city-scale data, interconnecting the different Fog areas and enforcing city-scale policies to be put in practice.

The above design paradigm can provide the necessary abstraction and virtualization for a next-generation C-ITS system. Hiding the different heterogeneous details of the various RATs and CAVs and simplifying the deployment of a new ITS service. This system architecture can accommodate next-generation ITS applications with strict QoS constraints bringing us closer to the paradigm of MaaS. In the following sections, we will describe the core network architecture and system components.

\section{Candidate RATs for V2X links}\label{sec.comms}
Currently, the IEEE 802.11p/Dedicated Short Range Communications (DSRC)~\cite{R4} and its extension IEEE 802.11px~\cite{ieee802_11px},  3GPP's LTE-Advanced (LTE-A) Pro~\cite{lte_advanced_pro} with its  Cellular-V2X (C-V2X) capabilities~\cite{c_v2x} and, millimeter-wave (mmWave) frequency systems~\cite{IEEE802_11ad} have emerged as potential RATs for V2X communications. This section will briefly present the pros and cons of these technologies and their impact on vehicular communications. 
Table~\ref{tab:technology_comparison} summarizes the main features and the additional capabilities provided by all the considered communication solutions.


\begin{table*}[t]
\renewcommand{\arraystretch}{1.3}
\centering
    \caption{Candidate RAT solutions for V2X communications}
\begin{tabular}{|M{2.8cm}|M{3cm}|M{3cm}|M{3.6cm}|M{3cm}|}
\hline
\diagbox[innerwidth=2.8cm]{\textit{Feature}}{\textit{RAT}} & \textbf{IEEE 802.11p/DSRC}~\cite{802_11p_table} & \textbf{IEEE 802.11px}~\cite{ieee802_11px} & \textbf{C-V2X (LTE-A Pro)}~\cite{c_v2x},~\cite{latency_lte} & \textbf{mmWave (IEEE~802.11ad)}~\cite{80211ad_table} \\ \hline \hline
	Frequency Band & \SI{5.85}{\giga\hertz} - \SI{5.925}{\giga\hertz} & \SI{5.85}{\giga\hertz} - \SI{5.925}{\giga\hertz} & \SI{450}{\mega\hertz} - \SI{4.99} {\giga\hertz} \SI{5.725}{\giga\hertz} - \SI{5.765}{\giga\hertz} & \SI{57.05}{\giga\hertz} - \SI{64}{\giga\hertz} \\ \hline
	Channel Bandwidth & \SI{10}{\mega\hertz} & \SI{10}{\mega\hertz} &   Up to \SI{640}{\mega\hertz} & \SI{2.16}{\giga\hertz} \\ \hline 
	Range & $\leq$ \SI{1}{\kilo\meter} & $\leq$ \SI{1}{\kilo\meter} & $\leq$ \SI{30}{\kilo\meter} & $\leq$ \SI{50}{\meter} \\ \hline
	Bit Rate & \SI{3}{\Mbps}-\SI{27}{\Mbps} & Up to \SI{60}{\Mbps} & Up to \SI{3}{\Gbps} & Up to \SI{7}{\Gbps} \\ \hline
	End-to-End Latency & $\leq \SI{10}{\milli\second}$ & $\leq \SI{10}{\milli\second}$ & \SI{30}{\milli\second}-\SI{50}{\milli\second} (UL/DL) \SI{20}{\milli\second}-\SI{80}{\milli\second} (V2V) & $\leq \SI{10}{\milli\second}$ \\ \hline
    Link Establishment Latency & \textasciitilde\SI{0}{\milli\second} & \textasciitilde\SI{0}{\milli\second} & \SI{40}{\milli\second}-\SI{110}{\milli\second} & \SI{10}{\milli\second}-\SI{20}{\milli\second} \\ \hline
	Coverage & Intermittent & Intermittent & Ubiquitous & Intermittent \\ \hline 
	Mobility Support & $\leq$ \SI{130}{\kilo\meter\per\hour} & Under Investigation & $\leq$ \SI{350}{\kilo\meter\per\hour} & $\leq$ \SI{100}{\kilo\meter\per\hour} \\ \hline
	QoS Support & Yes & Yes & Yes & Yes \\ \hline 
	Broadcast Support & Yes & Yes & Yes & No \\ \hline
	V2I Support & Yes & Yes  & Yes & Yes \\ \hline
	V2V Support & Yes & Yes & Over PC5 Interface & Yes \\ \hline
	Relay Mode & Yes & Yes & Yes & Yes \\ \hline
	MIMO & No & Yes & Yes & Yes \\ \hline
	\end{tabular}
    \label{tab:technology_comparison}
\end{table*}

\subsection{IEEE 802.11p/DSRC and IEEE 802.11px}
IEEE 802.11p/DSRC represents the suite of IEEE 802.11s and IEEE P1609.x standards, describing a system operating in the frequency range \SI{5.850}{\giga\hertz} to \SI{5.925}{\giga\hertz}, with a decentralized architecture that supports V2X communications. The IEEE 802.11p  standard defines the PHY and MAC layers.
More specifically, on top of the MAC layer, 
a Carrier Sense Multiple Access with Collision Avoidance (CSMA/CA) access to the medium, supports different QoS profiles by the Enhanced Distributed Channel Access (EDCA) protocol. The IEEE 802.11p/DSRC implementation of the EDCA is inherited from IEEE 802.11e with little modification. With regards to the PHY layer, the Orthogonal Frequency Division Multiplexing (OFDM) mechanism is adopted, which allows users to achieve a maximum PHY transmission rate of \SI{27}{\Mbps} at a speed of \SI{130}{\kilo\meter\per\hour}. However, due to the overhead of the communication protocols, the actual network throughput achieved is limited to about \SI{15}{\Mbps}~\cite{vnc_paper}.
The reserved \SI{75}{\mega\hertz} of the spectrum is divided into seven channels each of \SI{10}{\mega\hertz} bandwidth, where channel 178 (the Control Channel, CCH) is solely intended for broadcasting safety and mission-critical messages, and six channels (the Service Channels, SCHs) being used for all other applications.

\begin{figure}[t]
\centering
\includegraphics[width=1\columnwidth]{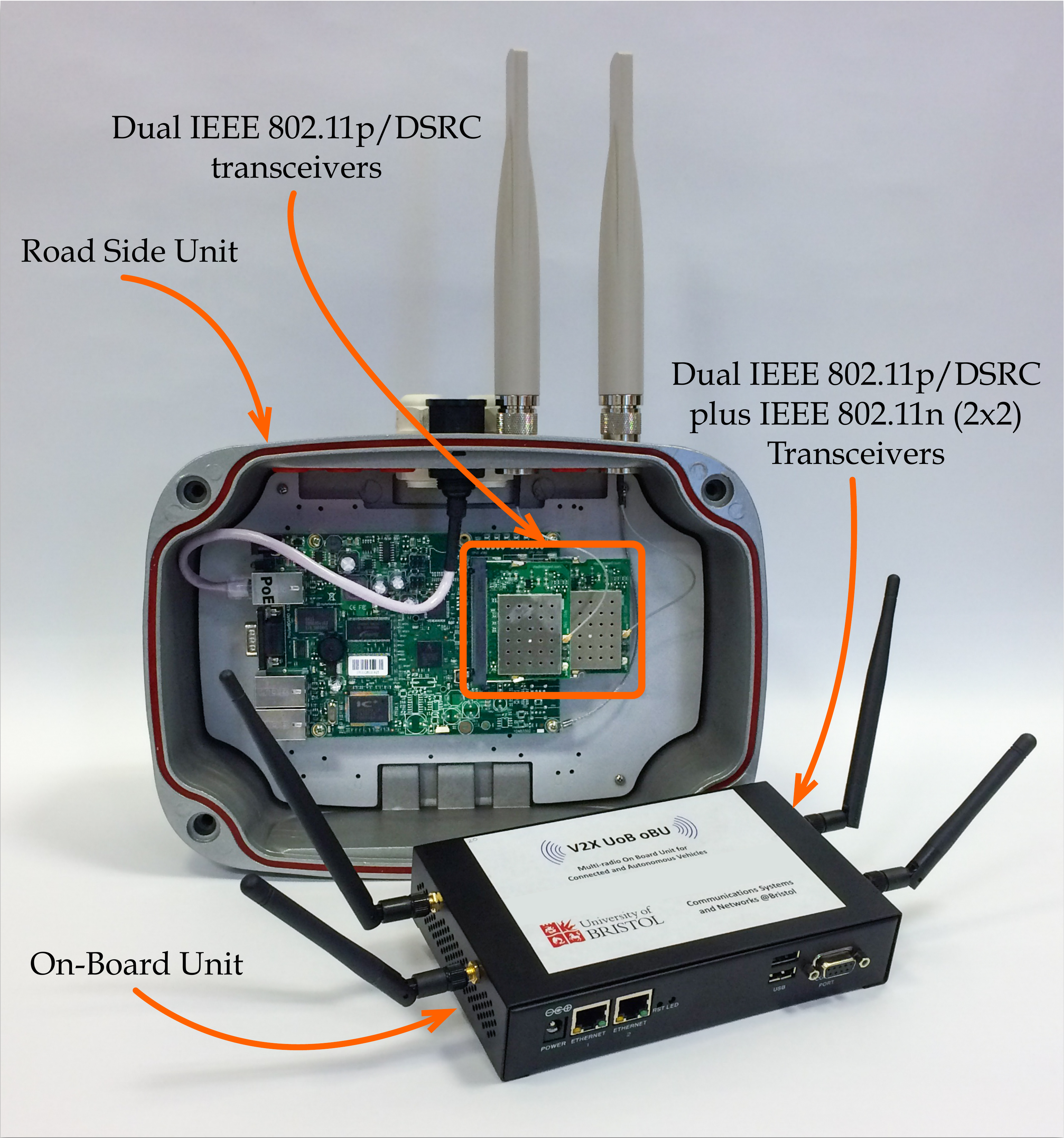}
\caption{Low-latency Linux Kernel implementation of IEEE 802.11p/DSRC units prototyped by the University of Bristol. The figure shows units designed for a road side and an on-board vehicle deployment.}
\label{fig:units}
\end{figure}

Due to the adopted MAC contention mechanism, the density of vehicles per area may have a disruptive impact on the overall end-to-end delay. This is caused by the lack of coordination among the devices, i.e., the number of channel erasures increases more than linearly with the vehicle density. As a result, IEEE 802.11p/DSRC tends to be more suitable for transferring low bitrate data streams in vehicular environments characterized by low-to-medium road density and vehicle speed. Examples of an IEEE 802.11p/DSRC Road Side Unit (RSU) and On-Board Unit (OBU) are shown in Fig.~\ref{fig:units}.

IEEE 802.11p/DSRC drawbacks are currently being addressed by an enhanced vehicular communication standard, denoted as IEEE 802.11px~\cite{ieee802_11px}. IEEE 802.11px builds the IEEE 802.11ac PHY layer. As such, the Low Density Parity Check (LDPC) codes for channel coding and the Space Time Block Coding (STBC) increase the performance under noisy channel transmissions improving the perceived Signal-to-Noise-plus-Interference (SINR). It is expected the packet delivery rate (PDR) to improve by $40\%$ at a transmitter-receiver distance of \SI{300}{\meter} compared to the legacy IEEE 802.11p. Also, Multiple-Input Multiple-Output (MIMO) antenna capabilities, the improved OFDM pilot layouts and the introduction of Very High Throughput (VHT) frame, are expected to enhance the channel capacity of a factor of $10$ compared to the IEEE 802.11p PHY layer. The new IEEE 802.11px/DSRC will adopt the Outside the Context of a BSS (OCB) operation of IEEE 802.11p/DSRC introducing backward compatibility with its predecessor. Despite the IEEE 802.11px being still in its early development stages, it is expected to replace the legacy IEEE 802.11p in the next years. It is worth noting that even though IEEE 802.11px is expected to improve the system performance compared to its predecessor, still it cannot accommodate the huge amount of data required by next-generation C-ITS services.

\subsection{3GPP LTE-Advanced Pro and C-V2X}
LTE and its major enhancement, LTE-A Pro, represent the 5G of cellular communications standards defined by 3GPP to provide high data rate, ubiquitous coverage and global connectivity for mobile cellular users~\cite{c_v2x}. 
The air interface can support Time-Division Duplexing (TDD), Frequency-Division Duplexing (FDD), and Half-Duplex FDD, as well as scalable channel bandwidths (\SI{1.4}{\mega\hertz} - \SI{20}{\mega\hertz}). Furthermore, a maximum of $32$ component carriers can also be aggregated in \mbox{LTE-A} Pro, leading to a maximum aggregated bandwidth of \SI{640}{\mega\hertz}. Support for Licensed Assist Access (LAA), enhanced LAA and LTE WiFi Aggregation (LWA) is also added, which means that additional bandwidth can be made available by aggregating data together from regular LTE bands, the 5GHz LTE-unlicensed spectrum, and common WiFi networks. 

Downlink and uplink access technologies are based on Orthogonal Frequency Division Multiple Access (OFDMA) and Single Carrier Frequency Division Multiple Access (SC-FDMA), respectively, thus guaranteeing high flexibility and efficiency in frequency-time resource scheduling. Due to the advanced MIMO capabilities, significant spectral efficiency can be obtained, even at high mobility speeds and under dynamic propagation environments. 
Besides, high data rates are potentially supported, ranging from \SI{300}{\Mbps} and \SI{75}{\Mbps} in the LTE downlink and uplink respectively, to theoretically \SI{3}{\Gbps} in the case of the LTE-A Pro downlink~\cite{RR0}. Multicast/broadcast services are also fully supported employing the evolved Multimedia Broadcast and Multicast Service (eMBMS) -- thus enabling broadcasting of vehicular service messages in cross-traffic assistance applications~\cite{7849790}.

Along with the eMBMS framework, V2V capabilities are built upon an enhanced LTE Direct mode and allow any network entity to engage to device-to-device (D2D) communications. The C-V2X operation of LTE-A Pro introduces two new radio interfaces. The cellular interface (namely as \emph{Uu}) that supports V2I links and the PC5 interface, which is responsible for the V2V communications. The configuration of the sidelinks (V2V links) depends on the availability of cellular coverage. When it is present, control information is transmitted to CAVs using the Uu interface and V2V communications is established.
As for drawbacks we observe, that radio resources need to be allocated by each base station to which each CAV connects to. Currently, this key procedure is only accepted in principle but is not part of any LTE-A standard amendments.

Due to the flat architecture, applications demanding low-to-medium latency requirements can be supported without affecting the network scalability. In particular, up to \SI{80}{\milli\second} end-to-end delay can be reached over an LTE-A Pro network. 
However, \emph{tactile-like} latency requirements smaller than \SI{10}{\milli\second} cannot be fulfilled in the presence of higher cellular traffic load.
Furthermore, terminals in idle mode need to re-establish a connection with the base station (additional link establishment latency of up to \SI{110}{\milli\second}), thus spending additional time to reach to the connected state. This leads to severe performance degradation in safety-critical applications~\cite{lte_survey}.

\subsection{Millimeter Wave Systems}
Systems based on mmWave are expected to play a pivotal role in 5G cellular systems. A mmWave system operates in the spectrum between \SI{30}{\giga\hertz} and \SI{300}{\giga\hertz}. For what concerns the application domain of local area networking, the IEEE 802.11ad standard is gaining momentum~\cite{C_ITS,tvtAndrea}. 
In this standard, the carrier frequencies are spread around \SI{60}{\giga\hertz}, with a channelization of \SI{2.16}{\giga\hertz}. 
Also, any mmWave system imposes the adoption of large antenna arrays to achieve high array gains through beamforming techniques. The high array gains, along with large channel bandwidths, allow the system to achieve high data rates (typically several gigabits-per-second). 
IEEE 802.11ad ensures data rates higher than \SI{7}{\Gbps} and an end-to-end latency smaller than \SI{10}{\milli\second}.

From the signal propagation perspective, Line-of-Sight (LOS) communications are characterized by path loss exponents smaller than $2.8$, while Non-Line-of-Sight (NLOS) communications may present much higher path loss exponents. In fact, due to their reduced wavelength, mmWave systems are susceptible to blockages. For these reasons, typical NLOS path loss exponents span between $3.8$ and $5.6$~\cite{RR0}.

As a prospective 5G wireless solution, mmWave systems proved to be a viable alternative to traditional cellular networks and wireless backhauling systems, while the possibility of using this technology to support the communications in next-generation ITS systems is being extensively investigated. In particular, the European Commission is currently considering the possibility of supporting standardization activities of ITSs based on mmWave systems to be operated across a dedicated band spanning between \SI{63}{\giga\hertz} and \SI{64}{\giga\hertz}~\cite{C_ITS}.

With ideal propagation conditions, mmWave systems significantly outperform vehicular communication systems based on the IEEE 802.11/DSRC and LTE/LTE-A standards. Obviously, large values of penetration loss and errors in the alignment of the antenna beams will have a disruptive impact on the stability of mmWave links. Also, the legacy beamforming training of IEEE 802.11ad can add an additional latency of up to \SI{20}{\milli\second} in a system~\cite{vtcIoannis}.

\section{Proposed ITS Agent Design for Next-Generation CAVs}\label{sec.ITS}
This section describes the proposed ITS agent design for next-generation ITSs featuring CAVs through heterogeneous RATs, as shown in Fig.~\ref{fig.System}. With regards to Section~\ref{sec:DM}, we observe that the ITS agent interacts with the Decision-Making agent to achieve ITS service goals, for instance, by implementing long-term driving maneuvers (ITS Agent to Decision Making agent signaling) or adapting the service goals in accordance with detected car accidents (Decision Making agent to ITS agent signaling). 
A FO responsible for a specific fog area, and most importantly the embedded Decision-Making agent, can directly access all the generated data within this area. Applications and services requiring data from other city regions or Fog areas can access them via the cloud-services communication links.

Since IEEE 802.11p/DSRC access network does not rely on any core networks, this ensures low latencies V2X communications at the cost of reduced coverage. To this end, it appears natural to relay safety/mission-critical messages via an IEEE 802.11p/DSRC network when: i) they are relevant to surrounding vehicles, ii) are characterized by low data rate, and iii) tolerate a \emph{tactile}-like end-to-end latency.
For instance, this is the case of messages signaling that a vehicle triggered its emergency braking system. This information is likely to be relevant only to the immediate surrounding vehicles.
On the other hand, if vehicles get involved in an accident in the middle of an intersection, this can have a disruptive impact on all the traffic flow across a large area of the city. In that case, it is worth taking advantage of an LTE-A Pro network to notify the disruption across geographically larger areas. Whereas, the authorities via the mmWave communication infrastructure can gain access to the vehicle camcorders or LiDARs to assess the severity of the accident (see Fig.~\ref{fig:NW_T}).

As mentioned in Section~\ref{sec:DM}, the proposed system paradigm is based on the idea of \emph{data scalability}, and a data stream admission control and RAT selection, all being part of the ITS agent entity. For this system, one base layer and two enhancement layers will be considered. The next sections present the key ITS agent components and their fundamental operations.

\subsection{Service Plane}\label{subsec:SP}
\begin{figure*}[tb]
\centering
\subfloat[Structure of an ITS agent]{\label{fig:SA}
	\includegraphics[width=0.93\columnwidth]{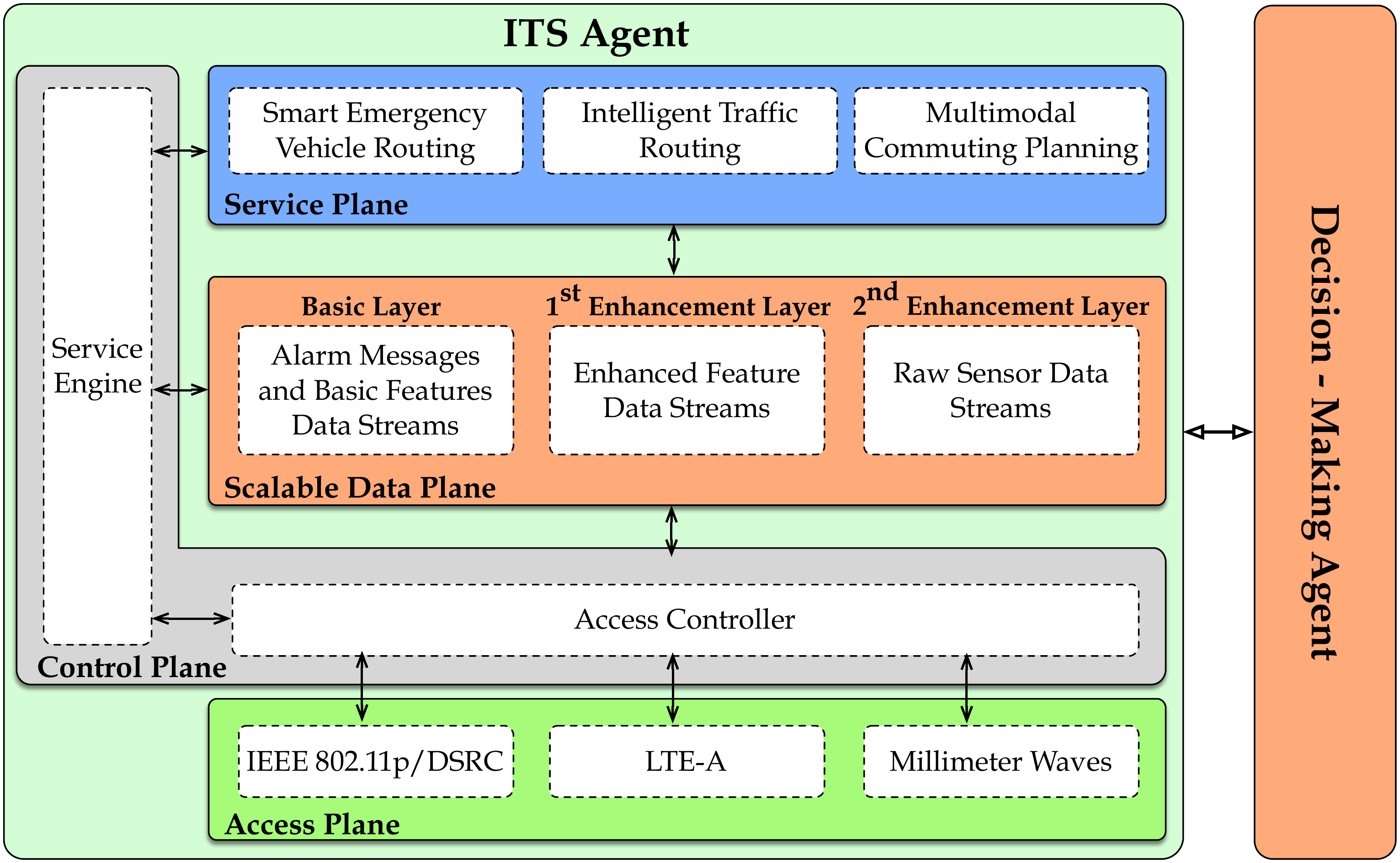}
}
\subfloat[Detail of the Access Plane]{\label{fig:SA_NW}
	\hspace{1cm}\includegraphics[width=1\columnwidth]{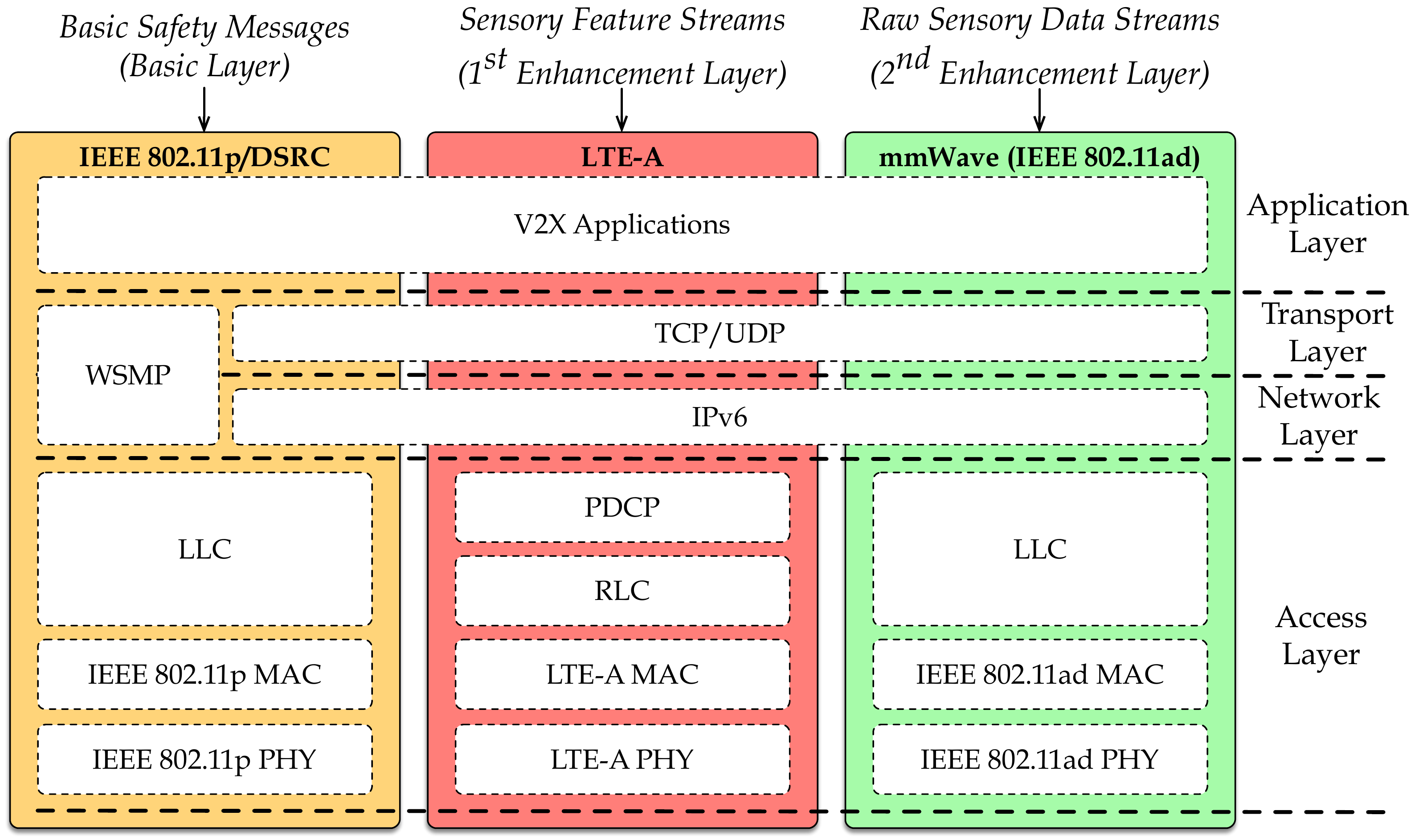}
}
\caption{Proposed ITS agent architecture and detail of the proposed Access Plane incorporating the considered RATs.}
\label{fig.System}
\end{figure*}

As illustrated in Fig.~\ref{fig:SA}, the Service Plane is the place where all the next-generation ITS services and applications are developed regardless of how the data streams will be mapped onto multiple sublayers and regardless of which RAT will be used to transmit each sublayer. We consider the following key next-generation ITS services, which are expected to enhance the CAV-based MaaS paradigm:
\begin{itemize}
    \item \textit{Intelligent Traffic Planning}: Future ITS traffic planning services are needed to reroute autonomous vehicles in the event of traffic jams, to coordinate traffic lights for offloading a congested road, or to provide drivers with essential up-to-date information.
    \item \textit{Smart Emergency Vehicle Routing}: This service is a specialization of the previous one, it assists emergency vehicles and provides the best route to reach a destination. Typical critical situations correspond to immediate medical assistance or disaster emergency management, such as fire alarms, where police vehicles need to promptly escort the fire trucks. Under these circumstances, the surrounding vehicles must be informed of the approaching emergency vehicles and rerouted to drastically reduce the traffic congestion. Furthermore, a traffic light synchronization system can minimize the overall journey time. A more complex use case scenario corresponds to a fully autonomous E-Ambulance system, where vehicles are equipped with health monitoring devices, such as wearable sensors, able to transmit the collected patience's data to the hospital or to a control center before reaching their destinations.
    \item \textit{Multimodal Commuting}: This service aims to dynamically and adaptively plan the route of a road user, combining different transportation systems, depending on start and end-points. For instance, an intelligent commuting system can notify drivers of available parking areas, where a shuttle bus service is offered to efficiently carry the employees to their workplace.
\end{itemize}

About the aforementioned ITS applications, Table~\ref{tab:key_messages} considers some examples of data streams being exchanged, their impact on the ITS service and the estimated amount of sensor data to be transmitted/received. As an example, it is worth mentioning that a state-of-the-art front looking radar can generate up to \SI{2800}{\Mbps}, while a camcorder operating at $640 \times 480$ pixels and $30$ frames per second, can generate more than \SI{400}{\Mbps} of uncompressed video data~\cite{nvidia}. As described in Section~\ref{sec:DM}, each ITS service has unique requirements with regard to the sensor data to be exchanged. The above figures of merit prove the necessity for a scalable data plane design and a flexible system architecture to accommodate these needs.

\begin{table*}[t]
\renewcommand{\arraystretch}{1.3}
\centering
    \caption{Relevant messages for next-generation ITS Services}
\begin{tabular}{|x{2cm}|M{6.5cm}|M{5cm}|M{2.5cm}|}
\hline
\textbf{ITS Services} & \textbf{Example Message Types} & \textbf{Impact on ITS Services} & \textbf{Estimated Sensor Data} \\ \hline \hline
\multirow{3}{2cm}[-1.6em]{\centering Intelligent Traffic Planning} & City-wide map grids and road shape reports. & Enabling the origin-to-destination long-term journey planning & \SI{10}{\Kbps} to \SI{10}{\Mbps} \\ \cline{2-4}
                          & City-wide knowledge of CAV positions & Congestion prevention & \SI{10}{\Kbps} to \SI{800}{\Kbps} \\ \cline{2-4}
                          & Routes and destination in low resolution & Congestion prediction and high level rerouting & \SI{80}{\Kbps} to \SI{800}{\Kbps}  \\ \hline
\multirow{3}{2cm}[-2em]{\centering Smart Emergency Vehicle Routing} & LiDAR sensor raw data streams exchanged and processed in real-time & Precise high-mobility maneuvers & \SI{50}{\Mbps} to \SI{250}{\Mbps} \\ \cline{2-4}
                          & Accurate representation of the nearby moving obstacles & Decision making through accurate object tracking & \SI{80}{\Kbps} to \SI{800}{\Kbps} \\ \cline{2-4}
                          & Trajectory paths with time profiles & Enhanced cooperation between CAVs improving long-term maneuver smoothness & \SI{80}{\Kbps} to \SI{800}{\Kbps}  \\ \hline
\multirow{3}{2cm}[-1.6em]{\centering Multimodal Commuting} & Available parking spaces in close proximity & Reducing the overall commuting time & \SI{10}{\Kbps} to \SI{10}{\Mbps} \\ \cline{2-4}
                          & City-wide knowledge of CAV positions & Refinement of the expected arrival time & \SI{10}{\Kbps} to \SI{800}{\Kbps} \\ \cline{2-4}
                          & Information on road disruptions (for e.g., accidents, adverse weather conditions) & Smart CAVs rerouting & \SI{30}{\Kbps} to \SI{100}{\Kbps}  \\ \hline
	\end{tabular}
    \label{tab:key_messages}
\end{table*}

\subsection{Scalable Data Plane}\label{subsec:SDP}
In the proposed system design, the Service Plane builds upon a Scalable Data Plane, which is in charge of:
\begin{itemize}
\item Handling data streams from the lower system planes and dispatching them towards the Service Plane;
\item Grouping data streams on the basis of their geographical relevance and QoS requirements.
\end{itemize}

The Scalable Data Plane assumes that each next-generation ITS service exchanges scalable data streams that comprise up to three independent data layers. The data streams mapped onto the base layer convey SAE J2735 messages for Road Safety Applications, in a V2X fashion. Among the fundamental SAE J2735 messages, it is worth mentioning the Basic Safety Messages (BSM), which are broadcast up to every \SI{0.1}{\second} and contain core vehicle information, such as vehicle size, GPS location, braking system status, etc. Along with the other SAE J2735 messages (e.g., Intersection Collision Avoidance, Road Side Alert, etc.), BSMs allow the proposed ITS design to support basic safety/mission-critical ITS functionalities, such as support to navigation, obstacle avoidance, traffic light status notification, etc. Given their low data rate, nearly \emph{tactile}-like latency, and local relevance, the V2X base layers are transmitted over IEEE 802.11p/DSRC communication links.

The considered next-generation ITS applications (see Section~\ref{subsec:SP}) build upon the aforementioned basic ITS functionalities and impact on potentially large areas of a city. These services are expected to deal with \emph{feature streams} generated by each vehicle out of its on-board sensors, which correspond to the V2X first enhancement layer in the proposed ITS design. An example of feature stream is given by the 3D bounding box representation of objects surrounding each vehicle~\cite{3d_objects}. These data are being processed using sensor data fusion techniques and independently generated by each CAV~\cite{3d_objects}. These highly refined information streams need to be shared among a large number of vehicles to allow them to take long-term decisions. In particular, the first enhancement layer requires a communication system providing large coverage and links capable of \mbox{megabits-per-second}, though there is no need for \emph{tactile}-like latencies. Hence, we refer to LTE-A Pro as the proposing RAT.

As discussed, the first enhancement layer is the result of sensory data processing carried out by each vehicle independently. Considering the Smart Emergency Vehicle Routing application, in the case of large-scale accidents, city-level emergency rooms may find it convenient to gather raw sensor data from multiple vehicles, combine them and then extract the required features. In fact, in the case of LiDAR data, combining raw data acquired from different locations can eliminate multiple blind spots and lead to a more accurate 3D bounding box object representation~\cite{Zhao2014165}. Streams of raw sensory data define the V2I second enhancement layer. Transmitting raw sensory data requires communication links capable of a gigabit-per-second, which implies the adoption of the mmWave infrastructure. However, due to the intermittent connectivity level associated with this technology, no reliability constraints should be associated with the second enhancement layer. 

\subsection{Access and Control Planes}
To guarantee high system flexibility and adaptability, we propose to separate the Access Plane, encapsulating all the considered RATs, from the Control Plane, which is responsible for the data stream admission control and RAT selection. As discussed in Section~\ref{sec:DM}, this approach is largely adopted in the research domain on SDN, where decoupling the network functionalities from the RAT in use enables high system programmability, and abstracts the network infrastructure from services~\cite{6834762}.

The Access Plane encompasses the three RAT solutions mentioned above, thus realizing a heterogeneous network capable of operating under various conditions and fulfilling diverse QoS profiles. As shown in Fig.~\ref{fig:SA_NW}, each of the adopted standards is characterized by different protocol stacks. Besides the PHY and MAC layer described in Section~\ref{sec.comms}, 
IEEE 802.11p/DSRC and IEEE 802.11ad mmWave include the typical Logical Link Control (LLC) layer in charge of flow and error control, whereas LTE-A adopts the Radio Link Control (RLC), responsible for error detection and recovery, and the Packet Data Convergence Protocol (PDCP) for packet integrity protection and header compression. 
Moreover, in addition to the traditional TCP/IP suite located on top of the access layer, IEEE 802.11p/DSRC embraces the Wireless Access in Vehicular Environments (WAVE) Short Message Protocol (WSMP), supporting the one-hop broadcast transmission of high priority and time sensitive data messages.

To achieve the necessary level of abstraction, we consider a Control Plane consisting of two distinct system components: the \emph{Service Engine} and the \emph{Access Controller}. 
The former is responsible for classifying the incoming data streams based on the Scalable Data Plane configuration, while a decision algorithm determines which data stream must be processed. The \emph{Access Controller} then chooses the appropriate RAT based on the decisions made by the \emph{Service Engine} in accordance with the Scalable Data Plane, as discussed in Section~\ref{subsec:SDP}, as well as taking into account the availability of the chosen technology. Both system components are integrated within the logical entity of the FO.

\subsection{Cooperation between Different RATs}
In the previous sections, we described the key components of the proposed multi-radio 5G architecture paradigm, and we established a connection between the different scalable data layers and the individual RATs. The abstraction and virtualization introduced by this design, as discussed, can pave the way for more reliable ITS applications and services. Building upon the flexibility provided by this system, we can enhance the performance of each RAT even further and overall, provide a more robust system. 

More specifically, the cooperation between different RATs can be enabled within a fog area. The FO can utilize the already exchanged ITS application data to fine-tune each radio.
For example, a significant fraction of the Beacon Intervals (BIs) introduced in IEEE 802.11ad is occupied by the necessary beamforming operations between the devices. In particular, as observed in~\cite{iet}, about $1/3$ of the BI length is to be allocated for beamforming operations. However, by taking advantage of the BSMs (exchanged via IEEE 801.11p/DSRC), which encapsulate position and heading information of a CAV, our FO unit can track a vehicle on the road, inform the mmWave RSU about the estimated position of the vehicle and provide the necessary and pre-calculated beamforming information. This has the potential of improving the network throughput~\cite{iet,vtcIoannis}. 

By following a similar approach, FOs can contribute to delivering a system with advance user-authentication functionalities~\cite{vtcGiovanni}. In fact, users' credentials can be generated and processed by each FOs rather than in a single centralized node. Similarly, FOs can also be used to implement advanced PHY layer security features, for instance, based on random network coding strategies~\cite{Amjad,7858628}.

\section{Conclusions and Future Directions}\label{sec.cl}
In this paper, we proposed a novel C-ITS design paradigm for CAVs, based on the concepts of multi-layer application data streaming and heterogeneous networking. We discussed the main components of CAVs and the key role of wireless connectivity for message exchanging and cooperation, as well as the candidate RATs for V2X communications. Finally, we presented the proposed C-ITS architecture, explored its main components and listed the benefits of our design paradigm.

Future research efforts will be targeted towards the performance benchmarking of the underlying heterogeneous access networks in city-wide network deployments. 

\bibliographystyle{icstnum}
\bibliography{bib.bib}

\end{document}